\documentclass[a4paper,11pt]{article}
\usepackage[utf8]{inputenc}
\usepackage{lmodern}
\usepackage[T1]{fontenc}
\usepackage{amsmath,amsthm,amsfonts,amssymb,bm}
\usepackage{dsfont}     
\usepackage{enumitem}
\usepackage{graphicx}
\usepackage{booktabs}
\usepackage{xcolor}
\usepackage{authblk}
\usepackage[authoryear]{natbib}
\usepackage[right=2.5cm,left=2.5cm,top=2cm,bottom=3cm]{geometry}
\usepackage[font={small,it}]{caption} 
\usepackage[colorlinks,linkcolor=blue,citecolor=blue,urlcolor=blue]{hyperref}
\usepackage{subfig}
\usepackage{mathtools}

\usepackage{multirow}
\usepackage{makecell} 
\usepackage{afterpage}
\newcommand\freefootnote[1]{%
  \let\thefootnote\relax%
  \footnotetext{#1}%
  \let\thefootnote\svthefootnote%
}

\newcommand{\SIGMA}{\bm{\Sigma}}

\newcommand{\THETA}{\bm{\Theta}}
\newcommand{\GAMMA}{\bm{\Gamma}}

\title{ \textsc{Model-based clustering for covariance matrices via penalized Wishart mixture models} }
\author[1]{Andrea Cappozzo}
\author[2]{Alessandro Casa}
\affil[1]{Department of Economics, Management and Quantitative Methods, University of Milan}
\affil[2]{Department of Economics, University of Bergamo}
\date{}                     

\begin{document}
\maketitle

\begin{abstract}
\freefootnote{Authors contributed equally to this work}Covariance matrices provide a valuable source of information about complex interactions and dependencies within the data. However, from a clustering perspective, this information has often been underutilized and overlooked. Indeed, commonly adopted distance-based approaches tend to rely primarily on mean levels to characterize and differentiate between groups. Recently, there have been promising efforts to cluster covariance matrices directly, thereby distinguishing groups solely based on the relationships between variables. From a model-based perspective, a probabilistic formalization has been provided by considering a mixture model with component densities following a Wishart distribution. Notwithstanding, this approach faces challenges when dealing with a large number of variables, as the number of parameters to be estimated increases quadratically. To address this issue, we propose a sparse Wishart mixture model, which assumes that the component scale matrices possess a cluster-dependent degree of sparsity. Model estimation is performed by maximizing a penalized log-likelihood, enforcing a covariance graphical lasso penalty on the component scale matrices. This penalty not only reduces the number of non-zero parameters, mitigating the challenges of high-dimensional settings, but also enhances the interpretability of results by emphasizing the most relevant relationships among variables. The proposed methodology is tested on both simulated and real data, demonstrating its ability to unravel the complexities of neuroimaging data and effectively cluster subjects based on the relational patterns among distinct brain regions.

\end{abstract}

\section{Introduction}\label{sec:introduction}
Covariance matrices play a pivotal role in many different scientific fields, as they encode linear relationships among observed variables, thus shedding light on how various complex and multidimensional phenomena are interrelated. In finance, they are essential for portfolio theory as they guide and inform investment strategies. In genomics, covariance matrices capture the dependencies among genes. As such, they are crucial for understanding genetic variation, playing a key role in identifying associations with traits and diseases, as well as in uncovering the underlying structure of complex genetic data. In neuroscience, they are often employed to investigate brain connectivity by examining correlations between different brain regions, providing insights into brain activities. Furthermore, covariance matrices serve as a cornerstone in many multivariate data analysis techniques such as, to mention a few, principal component analysis and factor analysis. Consequently, there is a substantial body of work in literature dedicated to the precise estimation of covariance matrices, especially in those challenging situations where the number of observed variables is closed or larger than the number of observations; readers can refer to \citet{pourahmadi2013high} and references therein for a detailed review on the topic. 

From a cluster analysis viewpoint, the informative content of the covariance matrices has often been somewhat overlooked. In fact, ubiquitous approaches, such as k-means and hierarchical clustering, primarily focus on differences on the mean levels when attempting to group multivariate observations. Nonetheless, recently some scattered attempts have been made to propose clustering procedures aiming to identify groups of covariance matrices, therefore partitioning statistical units based on the intrinsic informative content of these objects. For example, in the seminal paper by \citet{Dryden2009}, the statistical analysis of covariance matrices is thoroughly explored, leading to the introduction of non-Euclidean distances that are more appropriate in this context. Building on this, distance-based clustering approaches specifically thought for covariance matrices have been introduced \citep[see e.g.,][]{cabassi2018three,cappozzo2018object}. From a model-based perspective, \citet{hidot2010expectation} have devised an approach to provide a probabilistically grounded formalization to this problem. More specifically, they consider a mixture model in which the component densities are assumed to follow a Wishart distribution, making it naturally suited for modeling covariance matrices. To obtain practical estimates of the parameters and the resulting partition, the authors introduce an EM-algorithm specifically designed to handle the characteristics of the proposed model. Despite being reasonable and successfully tested on real-world applications, this approach shows some limitations in high-dimensional situations. In fact, as stated by the authors, when the dimensionality of the covariance matrices to be clustered increases, the proposed method may encounter numerical and computational instabilities, jeopardizing the effectiveness of the approach.

To address this limitation, in this work we introduce a sparse Wishart mixture model. Sparsity is induced in the component-specific scale matrices of the Wishart component densities by devising a penalized estimation procedure. More specifically, we develop a tailored EM-algorithm where covariance graphical lasso is embedded in the maximization step.  By obtaining sparse estimates, our approach helps to alleviate some of the challenges associated with covariance estimation when dealing with a large number of variables. On one hand, it reduces estimation instabilities, leading to more reliable estimates. On the other hand, it simplifies the interpretation of the results by shrinking some of the entries of the component scale matrices toward zero, allowing to focus on the most relevant relationships among variables. Moreover, it lends itself to a visual interpretation in terms of Gaussian covariance graph models \citep{chaudhuri2007estimation}. Our proposal is employed to cluster subjects based solely on the information provided by their functional networks, obtained via functional magnetic resonance imaging (fMRI), and encoding the relationships in the activity of different brain regions. By highlighting distinct groups, the procedure serves as a stepping stone to unveil complex patterns in the functional brain connectivity and in its relation with anatomical connections and with neurological diagnosis and diseases. Lastly, the induced sparsity allows to focus on relevant different relationships among brain regions across groups, thus facilitating the interpretation of the results.

The remainder of the paper is structured as follows. In Section \ref{sec:preliminaries}, we provide an overview on the model-based clustering framework, focusing specifically on approaches aiming to partition covariance matrices. In Section \ref{sec:sparseWish}, we motivate and present our proposal, covering both model estimation and selection. Sections \ref{sec:Simulations} and \ref{sec:realDataAnalysis} assess the performance and applicability of the proposed approach, testing it both on simulated and on real data. Lastly, Section \ref{sec:Conclusion} concludes the paper with some final remarks and discussing potential future research directions.

\section{Preliminaries and related work}\label{sec:preliminaries}
Model-based clustering \citep{fraley2002model,mcnicholas2016mixture,bouveyron2019model} provides a widely used probabilistic-based strategy to cluster analysis. This approach posits that data stem from a finite mixture distribution, describing the heterogeneity of the data and reflecting the presence of diverse sub-populations or clusters. Conventionally, from a frequentist perspective maximum likelihood estimation is performed by means of the EM-algorithm \citep{dempster1977maximum}, utilizing a data augmentation approach where latent group indicators are treated as missing data. From a practical standpoint, once the estimates of the parameters are obtained, each observation is assigned to the pertaining cluster according to the maximum a posteriori (MAP) rule, assuming a one-to-one correspondence between mixture components and groups. 

When continuous data are observed, in most of the cases Gaussian densities are considered as mixture components. However, there has also been a growing attention towards more flexible modelling choices, potentially able to accommodate non-elliptical and asymmetric cluster shapes; readers can refer for example to \citet{mclachlan1998robust,lin2009maximum,lin2010robust,andrews2011model,gollini2014mixture,browne2015mixture,tomarchio2024model} for relevant methodological proposals on the topic. Additionally, in recent years, it has become increasingly common to encounter unique data structures, introducing new challenges in the modeling phase. Consequently, innovative strategies have been developed to accommodate these frameworks, adapting model-based clustering to handle functional data \citep{bouveyron2011model,bouveyron2015discriminative}, network data \citep{snijders1997estimation,handcock2007model}, time-dependent data \citep{de2008model,mcnicholas2010model}, and matrix-variate data \citep{viroli2011finite}, among others.

Along the same line, \citet{hidot2010expectation} have recently proposed a Wishart mixture model, which allows to extend the model-based approach to partition cross-product matrices. Formally, let $\GAMMA = \{\GAMMA_1, \dots, \GAMMA_n \}$ be a sample of $n$ squared, symmetric and positive semi-definite matrices, with $\GAMMA_i \in \mathbb{R}^{p \times p}$, $i=1,\ldots,n$. These matrices clearly share a strong connection with sample covariance matrices, differing from them only by a normalization term; consequently, they encode linear relationships among the $p$ observed variables. Implicitly, it is assumed that $\GAMMA_i$ has been generated as the cross-product $\mathbf{X}_{(i)}^T\mathbf{X}_{(i)}$, where $\mathbf{X}_{(i)} = \{ \mathbf{x}_1^{(i)}, \dots, \mathbf{x}_{n_i}^{(i)}\}$ is a random sample with $\mathbf{x}_l^{(i)} \sim \mathcal{N}_p(\mathbf{0}, \SIGMA)$, for $l = 1, \dots, n_i$. According to this assumption, $\GAMMA_i$ is distributed as a central Wishart with scale matrix $\SIGMA$ and degrees of freedom $n_i$, respectively the covariance matrix and the size of the underlying sample $\mathbf{X}_{(i)}$. Note that, in the following, we assume to have observed directly $\GAMMA_i$, without having access to $\mathbf{X}_{(i)}$ and considering the sample size $n_i$ as unknown, for $i = 1, \dots, n$. To cluster the observed matrices, the authors posit that $\GAMMA_i$ arises from a Wishart mixture model, with density given by 
\begin{eqnarray}\label{eq:wishMixture}
    f(\GAMMA_i; \THETA) = \sum_{k = 1}^K \tau_k f_{\mathcal{W}}(\GAMMA_i; \SIGMA_k, \nu_k),
\end{eqnarray}
where $K$ is the number of mixture components, $\tau_k$'s are the mixing proportions with $\tau_k > 0, \forall k = 1, \dots, K$, $\sum_{k = 1}^K \tau_k = 1$, and $\THETA = \{\THETA_k \}_{k =1}^K$ with $\THETA_k = \{\tau_k, \SIGMA_k, \nu_k\}$ the set of model parameters. Lastly, $f_{\mathcal{W}}(\cdot; \SIGMA_k, \nu_k)$ denotes the density of a Wishart distribution with component specific scale matrix and degrees of freedom denoted with $\SIGMA_k$ and $\nu_k$ respectively, which reads as follows:
$$
f_{\mathcal{W}}(\GAMMA_i; \SIGMA_k, \nu_k) = \frac{\vert \GAMMA_i \vert^{\frac{\nu_k - p - 1}{2}}\text{exp}\left\{-\frac{1}{2}\text{tr}\left(\SIGMA_k^{-1}\GAMMA_i \right)\right\}}{2^{\frac{\nu_k p}{2}} \vert \SIGMA_k \vert^{\nu_k/2} \gamma_p \left(v_k/2 \right)},
$$
with $\gamma_p(\cdot)$ denoting the multivariate Gamma function:
$$\gamma_p(a)=\pi^{p(p-1)/4}\prod_{j=1}^p \gamma\left( a+(1-j)/2 \right),$$
where $\gamma\left( b \right) = \int_{0}^{+\infty} t^{b-1}e^{-t} dt, b > 0$, is the Gamma function.

In the work by \citet{hidot2010expectation} parameters are estimated by means of a tailored EM-algorithm and, coherently with the model-based clustering framework, the final partition of the matrices $\GAMMA_1, \dots, \GAMMA_n$ in $K$ clusters is obtained resorting to the MAP rule. The authors demonstrate the effectiveness of the clustering method, which implicitly groups the underlying unobserved samples \(\mathbf{X}_{(i)}\) based solely on the relationships among the observed variables, as encoded in the estimated group-specific scale matrices \(\SIGMA_k\).

While reasonable, this approach presents relevant drawbacks especially when dealing with scenarios where the ratio $p/n$ is relatively large. In fact, the authors show that the estimator for $\SIGMA_k$ is essentially the average of $\GAMMA_i$ matrices, weighted by their probability to belong to the $k$-th component. Nonetheless, when \( p/n > 1 \), the cross-product matrix, and consequently the sample covariance, becomes singular, failing to provide reliable insights into the relationships among the observed features. Even when the ratio \( p/n \) is smaller than 1 but still moderately large, the sample covariance matrix is known to be highly unstable. This instability arises because the matrix tends to accumulate numerous estimation errors due to the large number of free parameters involved \citep[see, e.g.,][for a thorough discussion]{pourahmadi2013high}. In this context, \(\hat{\SIGMA}_k\), for \( k = 1, \dots, K \), may not reliably characterize the relationships among the observed features, potentially compromising the soundness of the resulting partitions. Additionally, in a clustering framework, conducting post-hoc analyses to explore cluster-specific patterns is often of interest. These analyses can provide valuable insights and conclusions about the underlying population and the factors contributing to its heterogeneity. However, when \( p \) is large, interpreting the results becomes challenging, as it is difficult to discern which relationships among variables are driving the clustering.

For these reasons, in this work we extend the approach by \citet{hidot2010expectation} proposing a sparse mixture of Wishart distribution. The proposal builds upon the ever-expanding literature on sparse model-based clustering \citep[see][among the others]{pan2007penalized, zhou2009penalized,fop2019model,casa2022group} by extending its rationale to clustering covariance matrices. In a nutshell, our proposal addresses some of the aforementioned issues by providing sparse estimates of the matrices \(\SIGMA_k\) for \( k = 1, \dots, K \). On one hand, this yields more reliable results in high-dimensional settings, potentially refining the final partitions. On the other hand, it simplifies the interpretation of the results by capturing group-specific marginal dependency structures. The proposed method is based on a penalized likelihood framework, which is detailed in the following section.

\section{Sparse Wishart mixture models}\label{sec:sparseWish}
\subsection{Model specification}\label{sec:modSpec}

As previously highlighted, the method proposed by \citet{hidot2010expectation} has some weaknesses regarding its applicability in high-dimensional settings and its interpretability. To overcome these limitations, we propose an alternative method referred to as \textit{Sparsemixwishart}. Our method estimates parameters and subsequently performs clustering by maximizing a penalized log-likelihood defined as:
\begin{eqnarray}\label{eq:penLik}
    \ell_P(\THETA; \GAMMA) = \sum_{i = 1}^n \log \sum_{k = 1}^K \tau_k f_\mathcal{W}(\GAMMA_i; \SIGMA_k, \nu_k) - p_\lambda(\THETA)\, ,
\end{eqnarray}
where the first term is the log-likelihood of a mixture of Wishart distributions while the second one is a penalty term on the model parameters. As it is customary in the statistical learning with sparsity literature, $\lambda$ denotes the positive hyperparameter that controls the strength of the penalization.

Taking our steps from the covariance graphical lasso introduced by \citet{bien2011sparse}, in this work we consider the following penalty term
\begin{eqnarray}\label{eq:covGlasso}
    p_\lambda(\THETA) = \lambda\sum_{k = 1}^K \vert\vert \mathbf{P}*\SIGMA_k \vert\vert_1 \, ,
\end{eqnarray}
where $*$ denotes the element-wise multiplication and $\vert\vert \cdot \vert\vert_1$ is the $L_1$-norm with $\vert\vert A \vert\vert_1 = \sum_{jh} \lvert A_{jh} \lvert$. Lastly, $\mathbf{P}$ is a predefined matrix with non-negative entries, whose specifics will be discussed shortly.

The choice of the penalty \eqref{eq:covGlasso} and the geometry of the $L_1$-norm induce a certain degree of sparsity by shrinking to zero some of the entries of the matrices $\SIGMA_k, k = 1, \dots, K$, with the amount of sparsity depending on $\lambda$. This allows to alleviate potential issues arising when estimating association matrices in large-dimensional scenarios. Sparse representations of the component scale matrices introduce a convenient connection with \emph{covariance graphs models} \citep{chaudhuri2007estimation}. Within this framework, the observed variables correspond to distinct nodes in the graph, with edges representing marginal dependencies between them. Consequently, two variables are not connected by an edge when they are marginally independent. This not only facilitates convenient graphical visualizations of the results, but also simplifies interpretation by enabling a better characterization of the obtained partition by exploring how marginal dependencies among features vary across clusters.

The inclusion of the matrix \(\mathbf{P}\) in Equation \eqref{eq:covGlasso} has the potential to enhance the flexibility of the method, as clever specifications may allow for the inclusion of user-defined constraints or prior beliefs on variable relationships. Some guidance on this choice has been provided in the seminal paper by \citet{bien2011sparse}. Common approaches include using all-ones matrices with zeros on the diagonal to avoid shrinking its entries, or defining \(\mathbf{P}\) as an adjacency matrix with predefined patterns. For example, in neuroscience, the amount of white matter fibers connecting different brain regions can offer prior insights into the association structure within functional Magnetic Resonance Imaging; readers can refer to Section \ref{sec:realDataAnalysis} for further details.


\subsection{Model estimation}\label{sec:modEst}

Considering for the moment $K$ and $\lambda$ as fixed, estimates of the parameters are obtained by maximizing \eqref{eq:penLik} with respect to $\THETA$. In this work, we rely on a EM-algorithm specifically tailored for maximum penalized likelihood estimation. To this wise, we firstly define the \textit{penalized complete-data log-likelihood} related to (\ref{eq:penLik}) as
\begin{equation}\label{eq:complLik_pen}
    \ell_C(\THETA) =  \sum_{i =1}^n\sum_{k = 1}^K z_{ik}\log\tau_k f_{\mathcal{W}}(\GAMMA_i; \SIGMA_k, \nu_k) -  \lambda\sum_{k = 1}^K \Vert \SIGMA_k \Vert_1 \, ,
\end{equation}
where, without loss of generality, the dependence from $\mathbf{P}$ has been omitted to simplify the notation. 
As usual, $\mathbf{z}_{i} = (z_{i1}, \dots, z_{iK})$ is the realization of the latent group membership indicator variable $\mathbf{Z}_i$, with $z_{ik} = 1$ if the covariance matrix $\GAMMA_i$ belongs to the $k$-th cluster and zero otherwise. 

During each expectation step (E-step), the posterior probability of \(\mathbf{Z}_{i}\) is updated, allowing for the computation of the conditional expectation of equation \eqref{eq:complLik_pen}, commonly referred to as the Q-function. Subsequently, the Q-function serves as the objective function to be maximized in the maximization step (M-step) to obtain updated parameter estimates. A detailed description of the algorithm follows in the next subsections.

\subsubsection{E-step}\label{sec:estep}
At the $t$-th iteration, in the E-step the posterior probability of $\GAMMA_i$ to belong to the $k$-th component is computed conditionally on the parameter estimates $\hat\THETA^{(t-1)}$ obtained at the previous iteration. More specifically, $\hat{z}^{(t)}_{ik}$ are ordinarily updated as
$$
\hat{z}^{(t)}_{ik} = \frac{\hat{\tau}^{(t-1)}_k f_{\mathcal{W}}\left(\GAMMA_i; \hat{\SIGMA}^{(t-1)}_k, \hat{\nu}^{(t-1)}_k\right)}{\sum_{l = 1}^K \hat{\tau}^{(t-1)}_l f_{\mathcal{W}}\left(\GAMMA_i; \hat{\SIGMA}^{(t-1)}_l, \hat{\nu}^{(t-1)}_l\right)} \, ,
$$
for $i = 1, \dots, n$ and $k = 1, \dots, K$.

\subsubsection{M-step}\label{sec:mstep}
In the M-step, the updates for the parameter estimates are obtained by maximizing the penalized Q-function, which in this context is defined as follows:
\begin{align}\label{eq:Qfunct}
    Q(\THETA_1, \dots, \THETA_K) =  \sum_{i =1}^n\sum_{k = 1}^K \hat{z}^{(t)}_{ik} \biggl[ & \log\tau_k - \frac{\nu_k p}{2}\log 2 - \frac{\nu_k}{2}\log \vert\SIGMA_k\vert - \log \gamma_p\left( \frac{\nu_k}{2}\right) + \nonumber \\
    & \left( \frac{\nu_k - p - 1}{2}\right)\log \vert\GAMMA_i \vert - \frac{\text{tr}(\SIGMA_k^{-1}\GAMMA_i)}{2}  \biggr] -  \lambda\sum_{k = 1}^K \Vert \SIGMA_k \Vert_1 \, . 
\end{align}
The simultaneous maximization of \eqref{eq:Qfunct} with respect to all the parameters is unfeasible. Consequently, we resort to a partial optimization strategy cycling over three different steps. Firstly, note that closed-form updates for $\tau_k$ at the $t$-th iteration are available and are given by 
\begin{eqnarray}\label{eq:updateMixtProp}
    \hat{\tau}^{(t)}_k = \frac{n_k}{n}, \hspace{0.8cm} \hat{n}_k^{(t)}  = \sum_{i=1}^n \hat{z}_{ik}^{(t)}
\end{eqnarray}
for $k = 1, \dots, K$. Subsequently, the update for the degrees of freedom is obtained by maximizing \eqref{eq:Qfunct} with respect to $\nu_k$. Following the rationale outlined in \citet{hidot2010expectation}, this corresponds to solving the following equation for each mixture component separately:
\begin{eqnarray}\label{eq:updateDoF}
    \sum_{i = 1}^n  \hat{z}^{(t)}_{ik}\log \bigg\vert \frac{\GAMMA_i (\hat\SIGMA_k^{(t-1)})^{-1}}{2} \bigg\vert = \sum_{i = 1}^n \hat{z}^{(t)}_{ik}\sum_{j = 1}^p \psi \left( \frac{\nu_k - j + 1}{2} \right),
\end{eqnarray}
where $\psi(\cdot)$ is the digamma function. No closed-form solution exists; nonetheless, \eqref{eq:updateDoF} defines a one-dimensional root-finding problem with a continuous, monotonic function in \(\nu_k\). It is therefore solved numerically to obtain \(\hat{\nu}_k^{(t)}\) for \(k = 1, \dots, K\).

Updates for component scale matrices are more involved and they differ from the ones in \citet{hidot2010expectation}, as our proposal aims to induce sparsity in $\SIGMA_k$. More specifically, it is easy to see that maximization of \eqref{eq:Qfunct} with respect to $\SIGMA_k$ boils down to minimize the following objective function: 
\begin{eqnarray}\label{eq:updateSparseSigma}
    Q(\SIGMA_k) = \log\lvert \SIGMA_k \lvert + \text{tr}\left( \SIGMA_k^{-1} \tilde{\mathbf{S}}_k \right) + \frac{2\lambda}{\hat{n}^{(t)}_k\hat{\nu}^{(t)}_k}\Vert \SIGMA_k \Vert_1,
\end{eqnarray}
where $\tilde{\mathbf{S}}_k = (\sum_{i = 1}^n \hat{z}_{ik}^{(t)}\GAMMA_i)/(\hat{n}^{(t)}_k\hat{\nu}^{(t)}_k)$. Minimization of \eqref{eq:updateSparseSigma} with respect to $\SIGMA_k$ is equivalent to a covariance graphical lasso problem with a modified penalty coefficient being equal to $2\lambda/(\hat{n}^{(t)}_k\hat{\nu}^{(t)}_k)$. In this work, we adopt the coordinate descent based approach proposed by \citet{wang2014coordinate}, as implemented in the \texttt{R} \citep{Rsoft} package \texttt{covglasso} \citep{covglassoMichael}. 

The assessment of overall convergence of the algorithm involves monitoring increases in the penalized log-likelihood throughout each complete iteration. More specifically, the implemented EM-algorithm is considered to have achieved convergence when 
$$
\big\vert \ell_P(\THETA^{(t)}; \GAMMA) -  \ell_P(\THETA^{(t-1)}; \GAMMA) \big\vert \le \varepsilon
$$
where in the analyses $\varepsilon$ has been set to $10^{-6}$. 

One final computational aspect worth discussing is the initialization process, which is crucial for every deterministic algorithm. While multiple random initializations are a consistently viable option, they come with a significant computational cost. Despite their undeniable effectiveness, this approach can become impractical when clustering high-dimensional covariance objects. To address this issue, we propose using hierarchical clustering with non-Euclidean Riemannian distance \citep{Dryden2009} as a measure of dissimilarity to provide an initial partition of the data. This distance-based initialization is not only computationally efficient but also applicable when exploring different numbers of mixture components, as discussed in the upcoming section on model selection.


The whole procedure has been implemented mainly using the \texttt{R} software, with some routines written in \texttt{C++} to minimize the total computational time. The code is freely available in the form of an \texttt{R} package at \url{https://github.com/AndreaCappozzo/sparsemixwishart}. 

\subsection{Model selection}\label{sec:modSel}
In the previous sections, the proposed methodology has been presented considering both the number of clusters $K$ and the hyperparameter $\lambda$ as fixed. Nonetheless, in practical applications they are unknown and they need to be defined in order to obtain a partition. In this work, we select those $K$ and $\lambda$ which maximize a modified version of the Bayesian Information Criterion \citep[BIC,][]{schwarz1978estimating} defined as 
\begin{eqnarray}\label{eq:BIC}
    \text{BIC} = 2\log L(\hat{\THETA}) - d_0\log(n) 
\end{eqnarray}
where $\log L(\hat{\THETA})$ is the log-likelihood evaluated in $\hat\THETA$. Here, $d_0$ denotes the number of parameters not shrunk to zero by the penalized estimation procedure outlined in Section \ref{sec:modEst}. The adequacy of the BIC as model selection criterion in the mixture modelling framework has been demonstrated from a theoretical standpoint \citep[see e.g.,][]{keribin2000consistent} and empirically proven in a plethora of clustering applications \citep[see][and references therein]{bouveyron2019model}. Lastly, note that the formulation in \eqref{eq:BIC} has already proven useful also to tune the amount of shrinkage \citep{zou2007degrees} and in the penalized model-based clustering framework \citep[see e.g.,][]{pan2007penalized}.

\section{Simulation study}\label{sec:Simulations}
\subsection{Experimental setup}\label{sec:expSetup}
\begin{figure}[t]
\centering
    \includegraphics[width=\linewidth]{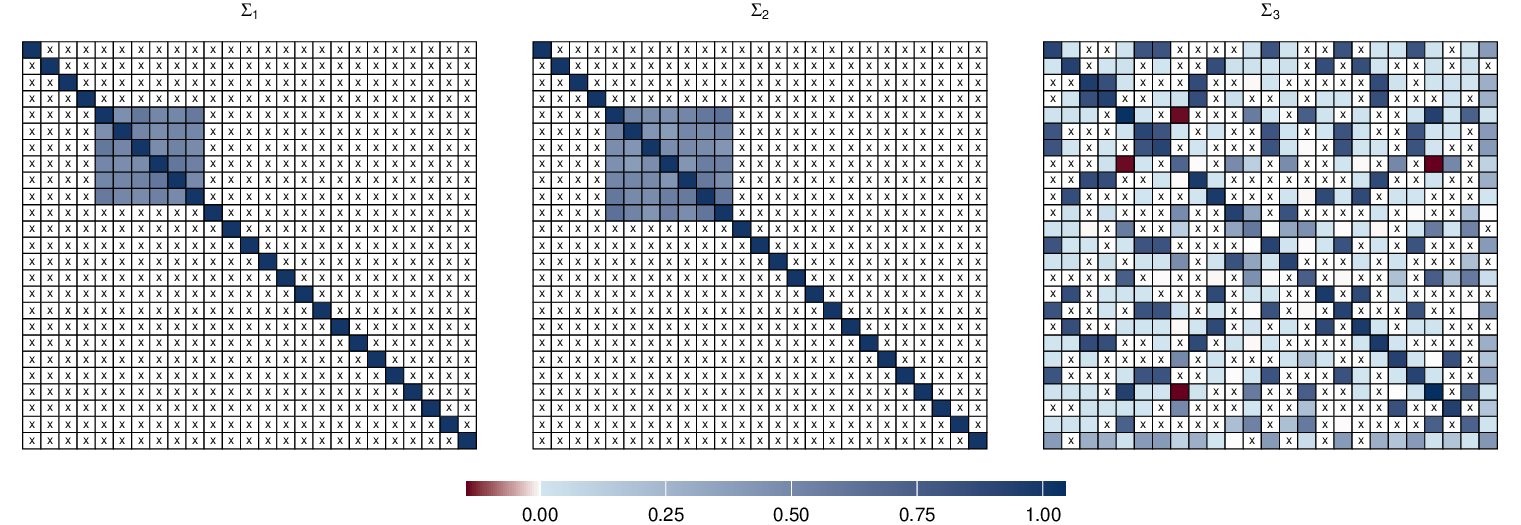}
    \caption{Heatmaps of the true $25\times 25$ scale matrices $\SIGMA_k$, $k=1,\,2,\,3$, considered in the simulated data experiment. A zero entry in the matrices is indicated with the symbol $\times$.}
  \label{fig:sim_true_sigma}
\end{figure}

In this section, we evaluate the performance of the proposed method on synthetic data, examining its capability to recover the underlying data partition while effectively identifying the sparsity structure in the component specific scale matrices. For each replication of the experiment, we draw $n=200$ random matrices with dimension $p \times p$, with $p=25$, from a Wishart mixture model with $K = 3$ components. We consider equal mixing proportions \(\tau_k = \frac{1}{3}\) for all \(k = 1, 2, 3\); with degrees of freedom of $30$, $30$, and $40$ for the first, second, and third components, respectively. The scale matrices are generated using two different mechanisms: an alternated-blocks structure for \(\SIGMA_1\) and \(\SIGMA_2\), and a sparse-at-random Erdős-Rényi graph structure \citep{erdHos1960evolution} for \(\SIGMA_3\). Such procedure results in component-wise different sparsity patterns and intensities, as displayed in Figure \ref{fig:sim_true_sigma}. We repeat the experiment $B = 100$ times and, alongside the methodology introduced in Section \ref{sec:sparseWish}, we consider the following competing models:
\begin{itemize}
\item \textit{Hidot Saint-Jean:} the original Wishart mixture model presented in \cite{hidot2010expectation}, where maximum likelihood estimation is used for model fitting, resulting in non-sparse estimates of the scale matrices \(\SIGMA_k\) for \(k = 1,\,2,\,3\).
\item \textit{Hclust-Euclidean}: hierarchical clustering using Frobenius distance as the measure of dissimilarity between random matrices, employing Ward's linkage method \citep{Murtagh2014}.
\item \textit{Hclust-Riemannian}: hierarchical clustering using Non-Euclidean Riemannian distance \citep{Dryden2009} as the measure of dissimilarity between random matrices, employing Ward's linkage method.
\end{itemize}
The simulation study has three main objectives. First, we aim to evaluate the ability to recover the true underlying group structure. To this end, all competing clustering methods are assessed using the adjusted Rand index \citep[ARI,][]{hubert1985comparing}. Secondly, we validate the quality of the estimates for the scale matrices \(\SIGMA_k\). Specifically, we compute the Frobenius distances $\Vert \SIGMA_k - \hat{\SIGMA}_k \Vert_F$ for the model-based methods only, namely our proposal \textit{Sparsemixwishart} and the original \textit{Hidot Saint-Jean} procedure. Lastly, we study the support recovery performance of our approach by determining whether the penalized estimation introduced in \textit{Sparsemixwishart} can accurately identify the true sparsity patterns in $\SIGMA_k$. To do so, we consider the $F_1$ score defined as follows:
\begin{equation}\label{eq:F1}
F_1 = \frac{\texttt{tp}}{\texttt{tp} + 0.5(\texttt{fp} + \texttt{fn})}.
\end{equation}
In Equation \eqref{eq:F1}, \texttt{tp} denotes the number of entries different from zero in $\SIGMA_k$ which are correctly estimated as such; \texttt{fp} represents the number of zero entries incorrectly identified as non-zero and \texttt{fn} is the number of non-zero elements which are wrongly shrunk to zero. 

For tuning the hyperparameter $\lambda$ a grid of $100$ equispaced elements is considered for each repetition of the simulated experiment, with the best shrinkage factor being selected using the modified BIC as suggested in Section \ref{sec:modEst}. Lastly, the issue of matching the estimated clusters with the original mixture components is addressed using the \texttt{matchClasses} function from the \texttt{e1071 R} package \citep{Meyer2023}. Simulation results are reported in the next subsection. 

\subsection{Simulation Results}\label{sec:sim_results}
\begin{figure}[t]
\centering
    \includegraphics[width=\linewidth]{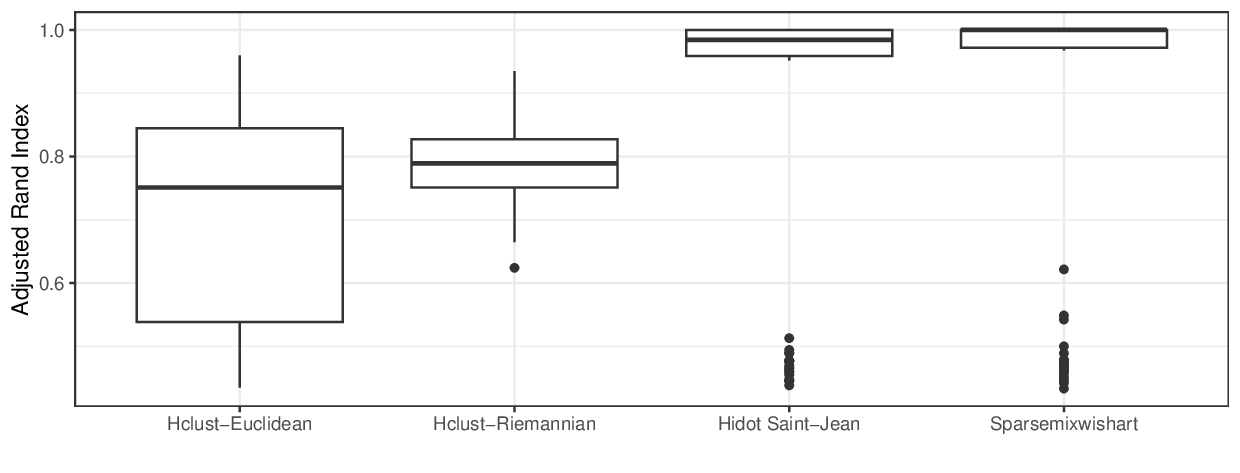}
    \caption{Boxplots of the ARI for the $B = 100$ repetitions of the simulated experiment. For the \textit{Sparsemixwishart} method, the shrinkage parameter $\lambda$ is selected according to the BIC defined in \eqref{eq:BIC}.}
  \label{fig:sim_ARI}
\end{figure}
\begin{figure}[t]
\centering
    \includegraphics[width=\linewidth]{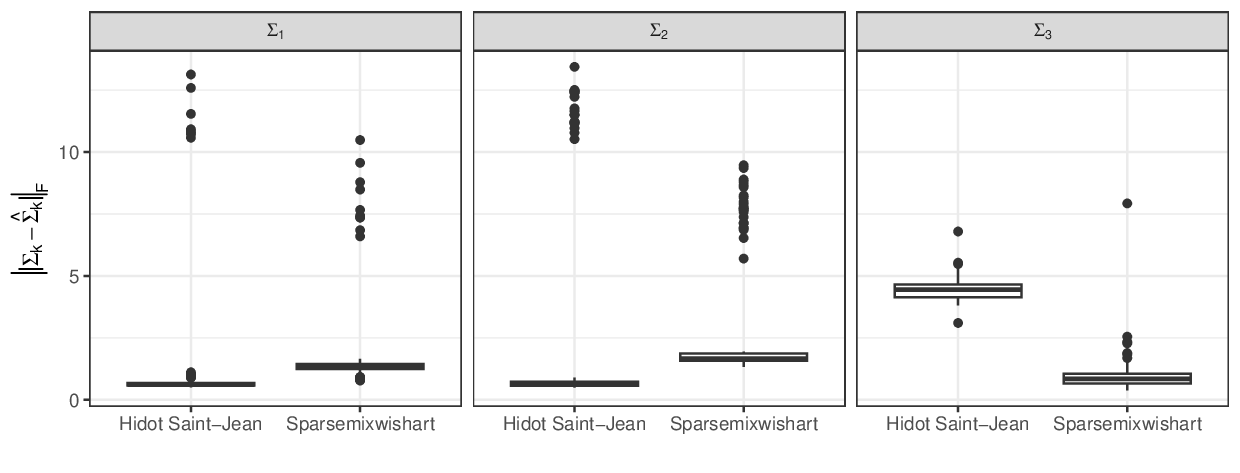}
    \caption{Boxplots of the Frobenius distance between true and estimated scale matrices $\SIGMA_k$, $k=1, 2, 3$, for $B=100$ repetitions of the simulated experiment.}
  \label{fig:sim_sparse_MSE_Sigma}
\end{figure}
\begin{figure}[t]
\centering
    \includegraphics[width=\linewidth]{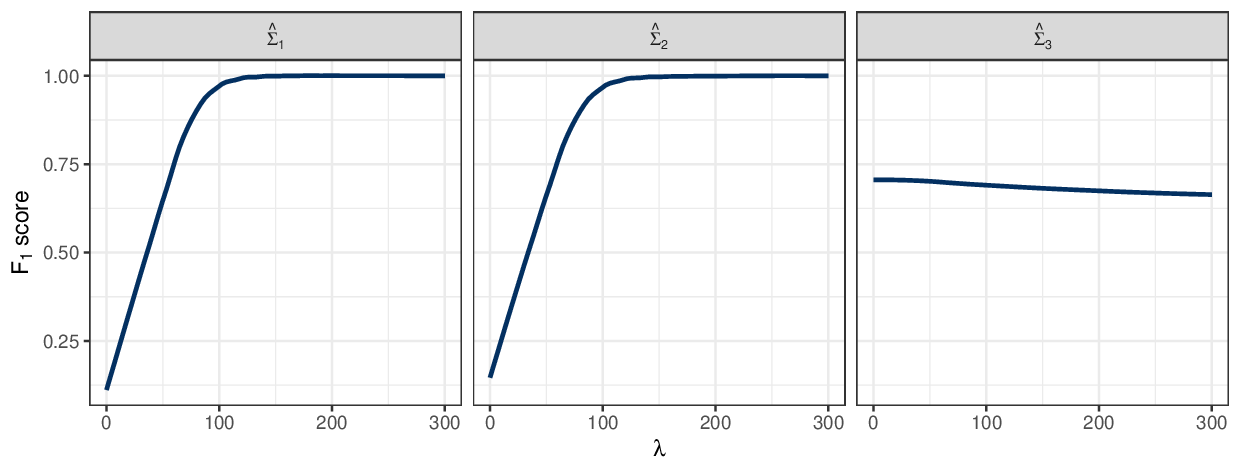}
    \caption{Smoothed lines plots of the component-wise different $F_1$ score for $B = 100$ repetitions of the simulated experiment, varying shrinkage factor $\lambda$.}
  \label{fig:sim_F1_smooth}
\end{figure}

Figure \ref{fig:sim_ARI} displays the empirical distribution of the ARI metric across the $B = 100$ repetitions of the simulated experiment for all the competing models introduced in the previous section. We immediately recognize that model-based methods outperform distance-based procedures in recovering the true underlying partition. It is noteworthy that hierarchical clustering based on Riemannian distance exhibits a higher median ARI and a lower standard error compared to its Frobenius counterpart. This advantage is likely due to the non-Euclidean nature of the space of positive semi-definite symmetric matrices, where a Riemannian metric may be more suitable for calculating distances between covariance objects \citep{Dryden2009}. Overall, our proposal demonstrates the best performance in terms of ARI, surpassing the already excellent results achieved by \cite{hidot2010expectation}. The reason for this improvement might stem from the covariance graphical lasso penalty introduced in \textit{Sparsemixwishart}, which allows for the sparse and more numerically stable estimation of the model parameters. In Figure \ref{fig:sim_sparse_MSE_Sigma}, we look at the Frobenius distances between true and estimated scale matrices $\SIGMA_k$, $k=1, 2, 3$. In detail, the $\Vert \SIGMA_k-\hat{\SIGMA}_k \Vert_F$ metric is, on average, roughly comparable for the first two components when using the \textit{Hidot Saint-Jean} method and our proposed approach, with the latter demonstrating lower variability. On the other hand, the novel \textit{Sparsemixwishart} methodology significantly outperforms, in terms of Frobenius distance, the unpenalized procedure when estimating \(\SIGMA_3\), ultimately leading to better results overall. After careful investigation, we reckon that this behavior can be attributed to the varying degrees of sparsity present in the true component scale matrices. A visual representation of this issue is displayed in Figure \ref{fig:sim_F1_smooth}, where we report smoothed line plots of the component-wise different $F_1$ scores computed as in \eqref{eq:F1}, as functions of the shrinkage factor $\lambda$. As the figure illustrates, while a higher penalty is more effective for the first two components in achieving better recovery of the sparsity structure, the impact of $\lambda$ on \(\SIGMA_3\) appears to be less influential. Note that, across the 100 replications of the simulated experiment, an average \(\lambda\) of 30.6 (SD = 10.8) was selected by the modified BIC in \eqref{eq:BIC}. This selection seems to favor a better structure recovery of the scale matrix \(\Sigma_3\), while potentially under-shrinking and inducing false positives in \(\Sigma_1\) and \(\Sigma_2\). Anyway, the visual inspection of Figure \ref{fig:sim_F1_smooth} suggests that an higher $\lambda$ might provide better overall $F_1$ scores across the three components. As such, the selection of $\lambda$ seems to be sub-optimal, if the recovery of the true structures of the scale matrices is the final aim. Needless to say the BIC, here used to select the shrinkage factor $\lambda$, is not specifically designed to maximize support recovery metrics. Therefore, when support recovery is the primary goal, a more appropriate criterion might be considered. For instance, in a different framework \citet{casa2024high} recently rephrase hyperparameter selection, needed to estimate sparse covariance matrices, in terms of a sequential hypothesis testing procedure, specifically aiming to recover the underlying structure. This might serve as a stepping stone for further generalizations to the covariance graphical lasso in the model-based clustering framework. Nonetheless, when dealing with mixture models, an additional complexity potentially influencing the results is given by the possibly different sparsity patterns and magnitudes across mixture components. Such difficulties are known in the penalized model-based clustering literature. As a matter of example, empirical results have demonstrated that an alternated-blocks structure is more easily recovered in this context \citep{fop2019model}. Recently, a solution has been proposed in \cite{casa2022group} in the framework of Gaussian mixture models. While adapting the methodology developed in \cite{casa2022group} to sparse mixtures of Wishart is beyond the scope of this paper, it could be a worthwhile avenue for future research.


Simulations show that the flexibility and parsimony offered by the covariance graphical lasso penalty not only improve the estimates compared to those obtained with the \textit{Hidot Saint-Jean} method in the presence of sparse scale matrices, but also enable better recovery of the true underlying partition in clustering covariance objects. These improvements are observed also in real data analyses in the neuroimaging context, as detailed in the next section.

 \section{Application to fMRI functional networks}\label{sec:realDataAnalysis}
The dataset analyzed in this work has been collected in a pilot study of the Enhanced Nathan Kline Institute-Rockland Sample project; a more detailed description of the aims of the project can be found at \url{http://fcon_1000.projects.nitrc.org/indi/enhanced/}. In details, brain imaging data are available for $24$ subjects over $70$ brain regions, constant across subjects and determined by the anatomical segmentation based on the Desikan atlas \citep{desikan2006automated}. More specifically, the study includes different sources of information: 
 \begin{figure}
\centering
    \includegraphics[width=\linewidth]{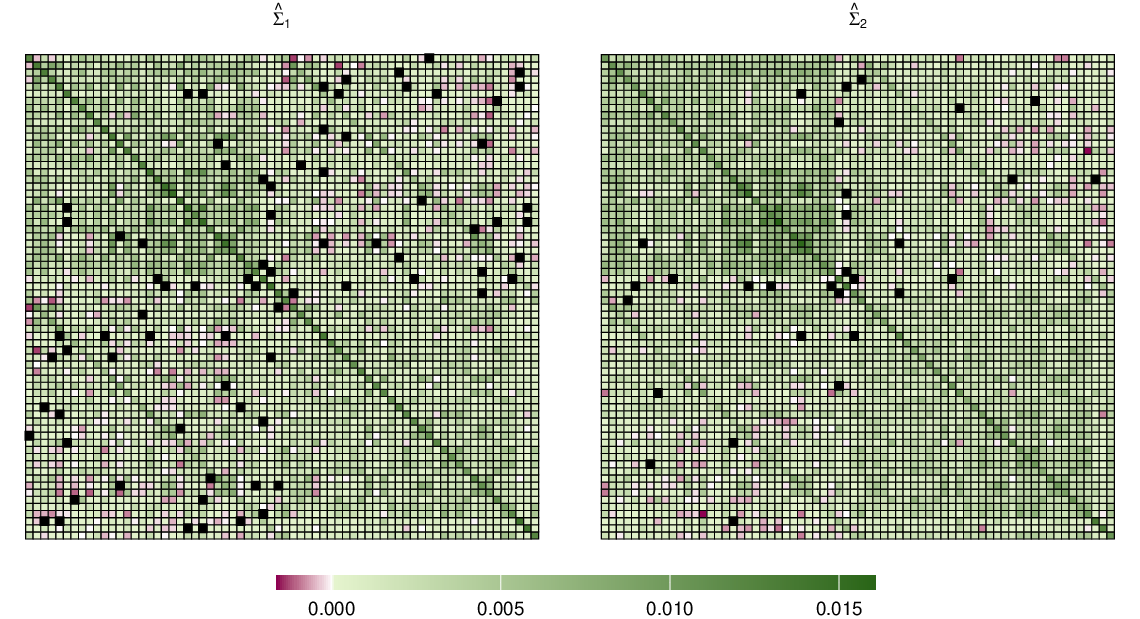}
    \caption{fMRI functional networks data. Estimated scale matrices $\hat{\SIGMA}_1$ and $\hat{\SIGMA}_2$ for the \textit{Sparsemixwishart} model, with $\mathbf{P}$ set to an all-ones matrix with zeros on the diagonal. Colors represent the values of the entries, with a $0$ entry denoted by the symbol $\blacksquare$.}
  \label{fig:fmri_sigma}
\end{figure}

\begin{itemize}
    \item \emph{Structural networks}, collected by means of Diffusion Tensor Imaging (DTI), measuring anatomical interconnections driven by the amount of white matter fibers connecting different brain regions.
    \item \emph{Dynamic functional activity}, collected by means of resting state fMRI (R-fMRI), measuring the activity of the $70$ brain regions over $404$ equally spaced time points, recorded when the subjects are not performing specific tasks.
    \item \emph{Functional networks}, representing the synchronization in brain activity between different areas of the brain, obtained computing the correlation of dynamic functional activity over time for the $70$ brain regions. 
\end{itemize}
Additionally, both subject-specific and region-specific covariates are available for each statistical unit, aiding in the interpretation of the results.

In this work, we focus mainly on the functional networks whose information is encoded in two arrays having dimension $70 \times 70 \times 24$, corresponding to two different scans performed on the same subjects. Given the way functional networks are constructed, they essentially encode information about how the activities of different brain regions are connected. Consequently, the goal of the analysis is to apply the developed methodology to cluster subjects based on the information contained in their functional networks. This approach not only facilitates the identification of distinct groups of subjects based on their brain activity patterns but also leverage the proposed penalized estimation scheme to emphasize the connections between regions that are most crucial for distinguishing different clusters. When paired with subject specific information, this might shed light on brain activity mechanisms and their relations with characteristics such as age, neurological diseases or previous diagnosis. Lastly, it is important to note that this application represents clustering in the truest sense, as we lack actual labels for direct comparison of results. Therefore, the clustering outcome will be interpreted and evaluated in conjunction with the available information about brain regions and subjects.

After conducting some preliminary analyses, we decided to focus exclusively on the first scan, being the second scan available only for a limited number of subjects. Additionally, two subjects were excluded due to the unavailability of their first-scan functional networks. Finally, two brain regions were discarded from consideration because they are labeled as \emph{unknown}, which hinders effective post-hoc interpretation of the results. Consequently, we consider for the analysis a sample of correlation matrices $ \{\GAMMA_i\}_{i=1}^n$, where $n = 22$ and $\GAMMA_i \in \mathbb{R}^{p \times p}$, with $p = 68$. The \textit{Sparsemixwishart} procedure outlined in Section \ref{sec:modSpec} was run for $K \in \{2,3,4\}$, letting the shrinkage parameters $\lambda$ varying within a pre-specified grid of values. Note that $\lambda = 0$ is also included in the grid, to allow for the comparison with the work by \citet{hidot2010expectation}.

Initially, $\mathbf{P}$ has been fixed to an all-one matrix with zero entries on the diagonal. In this scenario, the BIC in Equation \eqref{eq:BIC} selects a model with $K = 2$ components and a $\lambda$ value greater than zero, suggesting that our approach, which induces sparsity in the matrices $\SIGMA_1$ and $\SIGMA_2$, is appropriate in this context. Figure \ref{fig:fmri_sigma} presents the estimated scale matrices $\hat\SIGMA_1$ and $\hat\SIGMA_2$, with the resulting sparsity pattern clearly highlighted. We immediately notice that both matrices exhibit an alternating-blocks structure that mirrors the neuro-anatomic division of the brain into two hemispheres; with regions within the same hemisphere being more connected to each other than to regions in the opposite hemisphere. Secondly, despite the overall low level of estimated sparsity, the first scale matrix has more than three times as many entries shrunk to zero compared to the scale matrix of the second component. From a clustering perspective, this results in a partition of subjects into two groups, with sizes of 12 and 10, respectively. Analyzing additional subject information reveals significant differences between the groups. The average ages of the two clusters are 40 and 26.5, respectively. Additionally, the first group is predominantly right-handed, while the second group has a higher proportion of left-handed and ambidextrous individuals. Notably, the first group also has a higher number of subjects with a current diagnosis of mental disorder compared to the second group. Specifically, the diagnoses for the first group are mainly related to major depressive status. 
 \begin{figure}
\centering
    \includegraphics[width=\linewidth]{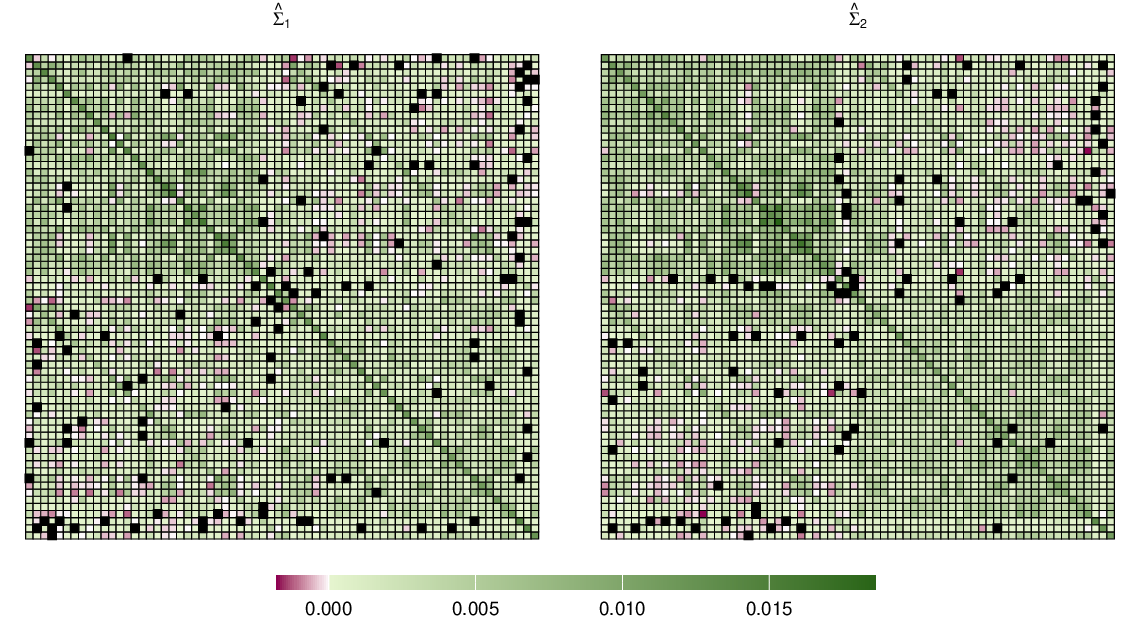}
 \caption{fMRI functional networks data. Estimated scale matrices $\hat{\SIGMA}_1$ and $\hat{\SIGMA}_2$ for the \textit{Sparsemixwishart} model, with a data-driven specification of $\mathbf{P}$ reflecting the structural anatomic interconnections among brain regions. Colors represent the values of the entries, with a $0$ entry denoted by the symbol $\blacksquare$.}
  \label{fig:fmri_StructNet_sigma}
\end{figure}

Alongside the analysis described in the previous paragraph, we applied the proposed methodology to the same data using instead a data-driven parameterization of the matrix $\mathbf{P}$. As mentioned in Section \ref{sec:modSpec}, clever specifications of this matrix allow to include additional available information on the phenomenon under study, thus enhancing the flexibility of the model. In the considered context, we leverage the knowledge of structural anatomic interconnections contained in the \textit{structural networks} dataset. In neuroscience, it is often assumed that brain activity reflects the underlying structural network and the connections of white matter fibers \citep{rykhlevskaia2008combining}. Based on this rationale, we devise a heuristic strategy to specify $\mathbf{P}$, ensuring that correlations between brain regions with fewer connections in terms of white matter fibers are more heavily penalized. Also in this scenario the BIC selects a model with $K = 2$ components and a positive shrinkage factor $\lambda$, providing further indications about the soundness of the proposal when compared to the one by \citet{hidot2010expectation}. The estimates $\hat\SIGMA_1$ and $\hat\SIGMA_2$ are reported in Figure \ref{fig:fmri_StructNet_sigma}. Compared to the results obtained with $\mathbf{P}$ set as an all-ones matrix with zeros on the diagonal, using a data-driven $\mathbf{P}$ results in a higher overall level of estimated sparsity. Although $\hat\SIGMA_1$ remains sparser than $\hat\SIGMA_2$, the difference in sparsity magnitude is smaller compared to the previous scenario, with the separation between the right and left hemispheres remaining clearly visible. From a clustering perspective the inclusion of the information on structural networks does not change the resulting partition. In fact, the resulting clustering aligns precisely with the previously discussed one, thereby preserving the insightful distinctions between the two groups in terms of age, handedness, and mental disorder diagnoses.

The proposed approach has demonstrated promising results in effectively separating subjects based on their functional networks while also incorporating external information related to structural anatomical interconnections. However, due to the very limited sample size, further investigation and more in-depth domain knowledge are needed to elucidate the determinants of specific mental disorder diagnoses and their connections to brain activity and the co-regulation of distinct brain regions.

\section{Discussion and conclusion}\label{sec:Conclusion}

As the complexity of routinely collected data continues to grow, effectively analyzing covariance matrices is becoming increasingly important. These objects play a crucial role in capturing intricate linear relationships among variables, making them essential in a variety of fields like finance, genomics, and neuroscience. When it comes to clustering, both distance-based and model-based methods can encounter challenges in high-dimensional settings, particularly when the sample size is limited.

To address this limitation, in the present paper we have introduced a novel sparse Wishart mixture model. By resorting to a penalized likelihood approach with a covariance graphical lasso penalty, we have enforced sparsity in the component-specific scale matrices. This solution not only reduces the number of parameters that need to be estimated but also enhances interpretability by shrinking negligible relationships among variables to zero. A tailored EM-algorithm to maximize the objective function has been developed, and specific strategies for initialization, convergence and model selection have been proposed.

A simulation study has been conducted to assess the effectiveness of our proposed method in clustering high-dimensional and sparse covariance objects. The results have demonstrated that our novel procedure outperforms state-of-the-art competitors in accurately detecting the underlying partition of the data. In addition, our approach has successfully recovered the true sparsity patterns within the component-specific scale matrices.

Finally, we have presented a pilot application where our procedure has been employed to cluster subjects based on their fMRI functional networks. The resulting groups have revealed distinct characteristics in terms of age, handedness, and mental disorder diagnoses. These promising results in such a complex application underscore the potential of our method to advance the field of neuroscience by providing clearer insights into functional brain connectivity and its relationship with various individual differences and clinical conditions.

Lastly note that sparse estimation of covariance matrices has attracted significant attention in recent years, especially in high-dimensional settings \citep{pourahmadi2013high, lam2020high}. In this work, we have focused on the covariance graphical lasso due to its seamless integration with the Wishart mixture model framework. However, exploring different penalization strategies or alternatives with different rationales such as thresholding \citep{bickel2008covariance,cai2011adaptive} or banding and tapering \citep{bien2016convex,bien2019graph} could further improve our methodology, as they represent promising avenues for investigation beyond the covariance graphical lasso approach. Additionally, future research could explore the incorporation of more flexible distributions for positive definite symmetric matrices. Potential candidates include the Riesz distribution \citep{Hassairi2001}, the inverse Riesz distribution \citep{Tounsi2012}, and the F-Riesz distribution \citep{Blasques2021}. These distributions offer alternative approaches to modeling covariance objects, which may enhance the adaptability and performance of clustering methods in complex and high-dimensional settings. Some of these options are currently under examination, and their potential contributions will be addressed in future research.

\section*{Acknowledgments}
We acknowledge Greg Kiar and Eric Bridgeford from NeuroData at Johns Hopkins University, who pre-processed the raw DTI and R-fMRI imaging data discussed in Section 5. Andrea Cappozzo acknowledges financial support within the ``Fund for Departments of Excellence academic funding'' provided by the Ministero dell'Università e della Ricerca (MUR), established by Stability Law, namely ``Legge di Stabilità n.232/2016, 2017'' - Project of the Department of Economics, Management, and Quantitative Methods, University of Milan. Lastly, the authors also express their gratitude to Elisa Borrini, whose master's thesis contains a preliminary version of this work.
\bibliographystyle{apalike}
\bibliography{biblio}

\end{document}